\documentstyle[12pt,version2,aps]{revtex}
\begin{document}

\renewcommand{\theequation}{\arabic{section}.\arabic{equation}}
\newcommand{\eqreset}{\setcounter{equation}{0}}
\setlength{\textheight}{20cm}
\vspace*{.9 in}
\begin{center}
{\large\bf FINE STRUCTURE AND FRACTIONAL {\bf M/N } AHARONOV-BOHM EFFECT}

\vspace{.5 in}
{\sc F.V. Kusmartsev$^{*1,2}$}

   and

 Minoru {\sc Takahashi $^2$}

 \vspace{.3 in}

{\it Department of Theoretical Physics}\\
{\it  University of Oulu}\\
{\it SF-90570 ~ Oulu, ~ Finland$^1$}\\

and

    {\it Institute for Solid State Physics,University of Tokyo}\\
        {\it Roppongi,Minato-ku,Tokyo 106 $^2$} \\

(received ~~~~~~~~~~~~~~~~~~  )
\end{center}

\vfill
\eject
\begin{abstract}

 We find a fine structure in the Aharonov-Bohm effect, which is
characterized by the appearence of a new type of periodic oscillations
having smaller fractional period and an amplitude, which may compare with
the amplitude of the conventional Aharonov-Bohm effect.
Specifically, at low density or strong coupling
 on a Hubbard ring can coexist along
 with the conventional Aaronov-Bohm  oscillations with the period equal to an
integer, measured in units of the elementary flux quantum, two additional
oscillations with periods $1/N$ and $M/N$. The integers $N$ and $M$ are the
particles number
and the number of down-spin particles, respectively.

 From a solution of the Bethe ansatz equations
 for $N$ electrons located on a ring in a magnetic field
 we show that
the fine structure is due to  electron-electron and  Zeeman interactions.
Our results are valid in the dilute density limit  and for an
 arbitrary
value of the Hubbard repulsion  $U$

 For the appearance or
 the existence of integer-fractional $M/N$ flux quantum
 periodic oscillations the key role is played by the Zeeman energy.
This is in an agrement with an assumption by Choy about importance of Zeeman
energy.
Without Zeeman energy there is a coexistence of the fractional
$1/N$ and half-flux quantum periodic oscillations, only.
With  increasing magnetic field, the Zeeman energy increases and
 the half-flux quantum period transforms  to the integer one.
This transformation is not  continious but via the appearance of
fractional $M/N$ oscillations, i.e.
 by some singular way.

The fractional $1/N$ regime occurs when the value $N< (LU/t)^{1/4}$, where $L$
is the number of sites.   The  coexistence regime  of integer-fractional
$M/N$ and fractional $1/N$ AB oscillations occurs
 when $ (LU/t)^{1/4} <<N<< (LU/t)$.
 The parity effect disappears in the fractional regime,
but returns in the integer-fractional regime.
We discuss the relation of the fine structure or the coexistence regime
 to existing experiments on
single rings and on array of rings, where integer and
 half flux quantum periods are, respectively, observed.
 The explanation for the fractional
$1/4$ AB  period, recently observed by Liu et al  is proposed.

\end{abstract}

\vfill
\eject

\section{Introduction}

In the limit
of large repulsion, when $U/t\rightarrow\infty$ Kusmartsev  \cite {Kus1}
found that the Aharonov-Bohm(AB) effect is fractional. In units of elementary
flux quantum  its period is equal to  $1/N$, where the $N$ is the
number of electrons on the ring.
This  effect is drastically different from conventional integer or
half-integer AB
effects\cite{Land},\cite{Kuli},\cite{Butt},\cite{Land1}\cite{Kus}.

This result has since been confirmed
in the investigation by Schofield, Wheatley and Xiang\cite {Scho}. The first
corrections were studied by Yu and Fowler
\cite{Yu}, where they first recognized the importance of
the magnon excitations for the Aharonov-Bohm effect.
 We extend their analysis and investigated the effects for the dilute system.
We find a new type of oscillations, related to a magnon excitation spectrum.
The new AB oscillations with  the period $M/N$ arising with the first
corrections
to the strong coupling limit or to the dilute density limit appears when
we take into account the polarized or partially polarized ring with $M$
down-spin electrons. That is
for the effect
it is very important  a role of the Zeeman interaction causing the polatization
effects. The importance of the Zeeman energy was noticed recently by
Choy \cite{Choy}.

We show that for the very dilute system, when the filling factor tends to zero,
as well as for strong-coupling limit, the fractional $1/N$
Aharonov-Bohm effect may coexist with  the integer conventional,
integer-fractional
or half-flux quantum periodic AB effect.  By the integer-fractional regime of
AB effect we mean the appearance of a new type of oscillations on the envelope
function of $1/N$ periodic oscillations, which display a new fractional
period. The period of these
oscillations is equal to $M/N$, where $M$ is the number of
down-spin particles. The $M/N$ oscillations
exist even when the $1/N$ oscillations are washed out.
 When the Zeeman interaction is neglected, i.e.
the Zeeman energy is zero and $M=N/2$,
there is a coexistence of the fractional $1/N$ and half-flux quantum periodic
oscillations. In contrast, when the system is almost polarized, i.e. $M=1$, the
fractional $1/N$ and integer periodic oscillations  coexist.

In previous work\cite{KWKT}  we found that finite scaling finite size effects
 were very important. There is a scaling behavior of the ground state energy
which  does not depend on the size $L$ or  on $U$, but depends only on
$N/UL=const=\alpha$,
where $U$ is measured in units of $t$ and $L$ is measured in units
of the lattice constant.  Such scaling occurs only at small values of $\alpha$.
Due to this scaling
symmetry  the fractional  Aharonov-Bohm effect may arise even for
small values of $U$, for the very dilute electron systems. In the present work,
we shall show that, such a symmetry is related to the structure
of the low energy excitations of the Hubbard Hamiltonian, namely,
with the spectrum of the spinon or magnon excitations.
In fact the parameter $\alpha$ defines the energy of the magnon excitations.
Thus, the shape of the new AB oscillations completely depends upon the
energy of the magnon excitations.

In fact, we get an amusing physical picture.
First, let us discuss the case $M=N/2$.
For an odd number of particles
 in the ground
state of the Hubbard ring, there already exists a spinon excitation.
The momentum of this excitation is associated with the top of the spin-wave
spectrum. This excitation is needed because of the parity effect,
discussed by Legget\cite{Legg}, Loss\cite{Loss} and Kusmartsev\cite{Kus},
for the case of interacting spinless fermions.

Thus, again with the first correction to the strong coupling limit
or to the dilute density limit we have shown that
on the Hubard ring there occurs the reconstruction of the parity effect,
reflecting
the structure of low-energy excitations of the Hubbard ring.
However, the parity effect obtained is distinct from the
parity effect for the mesoscopic ring with spinful electrons and disorder
\cite{Weis}
Inducing the flux the momentum of this spinon excitation, which exists
in the ground state, changes. This gives rise to the correspondence between
the spin excitation spectrum and the Aharonov-Bohm fine structure. With
the magnetic field the number of polarized electrons changes, i.e.
the spin excitation spectrum changes, this gives rise to the change
of the Aharonov-Bohm periodicity.

This occurs also for the general situation, with any number of particles on the
ring
and at different rational ratio $M/N$. In the latter case one may
see an additional fractional $M/N$ AB oscillations, which are due to
the structure of the spinon excitation spectrum;
the picture is as follows.
With the flux, the spinon changes its momentum, which is, in fact,
proportional to the flux. The change in the momentum does not
occur  continuously,
but via jumps related to the finite number $N$ of
the particles on the ring. Therefore these jumps are associated with the $1/N$
fraction of
elementary flux quantum. Thus, on the Hubbard ring with the flux
the spinons behave similar to spinless fermions, where the momentum
is also shifted with the flux, reconstructing the
 parity effects. The Fermi momentum of this fermions is related to the
number of polarized electrons $M$. This gives the new scale and the new
periodicity in Aharonov-Bohm effect, related to the ratio $M/N$.
However, the  fractional $M/N$ AB effects do not destroy
  the fractional $1/N$ effect associated with the bound state of $N$ particles.
Both effects may coexist, reflecting the separation of spin and charge
degrees of freedom.

\section{Main equations}

To demonstrate this picture we have studied the Hubbard Hamiltonian
\begin{equation}
H = - t \sum_{<i,j>, \sigma} a^\dagger_{i\sigma} a_{j\sigma}
     + U \sum^L_{i=1} n_{i+} n_{i-} \quad
\end{equation}
which is parametrized by the electron hopping integral $t$,
the on-site repulsive
Coulomb potential $U$ and the number of sites L.
The operator $a^\dagger_{i\sigma} (a_{i\sigma})$  creates (destroys)
an electron
with spin projection $\sigma$ ($\sigma = +\, \rm{or}\, {\bf - }$),
at the ring site $i$, and $n_{i\sigma}$
is the occupation number operator $a^\dagger_{i\sigma} a_{i\sigma}$. The
summation
in Eq. (1) extends over the ring sites $i$ or -- as indicated
by $<i,j>,\sigma$ -- over all distinct pairs of nearest-neighbor sites,
around the ring with the spin projection $\sigma$.
  For the case of the magnetic field we  use the same form
of the wave function as in  Refs \cite{LiebW},\cite{Suth} and
\cite {Kawa}, i.e.
\begin{equation}
\psi(x_1,\ldots,x_N)=\sum_P [Q,P]\exp[i \sum^N_{j=1} k_{Pj} x_{Qj}],
\end{equation}

where $P=(P_1,\ldots,P_N)$ and $Q=(Q_1,\ldots,Q_N)$
are two permutations of $(1,2,..., N)$, and N is the number of electrons.

     The   coefficients  $[Q,P]$,  as     well  as   $(k_1,...k_N)$,   are
determined  from the Bethe equations, which in a magnetic field  are
changed by the addition of the flux phase $2\pi f$ \cite{Suth},
\cite{Kawa},\cite {Staff}, \cite{Kus1},\cite{Fye}, \cite{Ferr}, and \cite{Sudb}
\begin{equation}
e^{i(k_j L-2\pi f)}=\prod^M_{\beta=1}
 \biggl({{it \sin k_j -i\lambda_{\beta}-U/4}
\over{{it \sin k_j -i\lambda_{\beta}+U/4}}} \biggr)
\end{equation}
and
\begin{equation}
-\prod^N_{j=1} \biggl({{it \sin k_j -i\lambda_{\alpha}-U/4}
\over{{it \sin k_j -i\lambda_{\alpha}+U/4}}} \biggr)=
\prod^M_{\beta=1}\biggl({{i\lambda_{\alpha} -i\lambda_{\beta}+U/2}
\over{i\lambda_{\alpha}} -i\lambda_{\beta}-U/2} \biggr)
\end{equation}
 $f$ being the flux in units of the elementary quantum flux $\phi_0$.
 The explicit form of the Bethe equations in a magnetic field is  \cite
{LiebW},\cite{Shas}, \cite{Kus1}, \cite{Fye},
\cite{Yu}
\begin{eqnarray}
Lk_j=2\pi I_j+2\pi f - \sum^M_{\beta=1} \theta(4(t\sin k_j -\lambda_{\beta})/U)
 \label{momenta}
\\
-\sum^N_{j=1} \theta(4(t\sin k_j -\lambda_{\beta})/U)=2\pi
J_{\beta}+\sum^M_{\lambda_{\alpha}=1}
\theta( 2(\lambda_\beta -\lambda_{\alpha})/U)\label{lambda}
\end{eqnarray}
where $\theta(x)=2 \arctan(x)$ and the quantum
numbers $I_j $ and $J_{\beta}$,
which are associated with the charge and spin degrees
of freedom, respectively, are either integers or hal- odd integers,
depending on the parities of the numbers of down and
 up-spin electrons, respectively, i.e.
\begin{equation}
I_j=\frac{M} {2} ~~~ {\rm (mod ~1) ~{and} }~~ J_{\beta}= \frac {N-M+1} {2}
{}~({\rm{mod} ~1).}
\end{equation}

\section{Expansion}

 In previous work\cite{KWKT} showned that
in the dilute density limit there exists a  small parameter $\alpha$, which
is
\begin{equation}
\alpha= \frac {N t} { L U} =\rho t/U
\label {parameter},
\end{equation}
where $\rho=N/L$ is
the filling factor.

{}From an expansion in $\alpha$ (
 $\alpha<<1$)
of the second Bethe equation (\ref {lambda}),  we get
the equations
\begin{eqnarray}
N\theta(t_{\beta}) -\frac{8} {U} \frac {1} {(1+t^2_\beta)} \sum^N_{j=1}
 \sin(k_j)=2\pi J_{\beta}+\sum^M_{\lambda_{\alpha}=1}
\theta( (t_\beta -t_{\alpha})/2),
\label{spin-lim-exp}
\end{eqnarray}
where $t_{\beta}=4 \lambda_{\beta}/U$.

Next step, we notice that the second term on
the left hand side  the eq.(\ref{spin-lim-exp}) is small and therefore
we may incorporate it into the argument $x$ of $\theta(x)-$, giving

\begin{eqnarray}
N\theta\biggl[t_{\beta} -\frac{4} {NU} \sum^N_{j=1}
 \sin(k_j)\biggr]=2\pi J_{\beta}+\sum^M_{\lambda_{\alpha}=1}
\theta( (t_\beta -t_{\alpha})/2),
\label{spin-lim-exp1}
\end{eqnarray}

With the substitution $x_\beta=t_\beta- \frac{4} {NU}\sum^N_{j=1}
 \sin(k_j)$, this equation is reduced to the Bethe equation
for an isotropic Heisenberg antiferromagnet on a ring having $N$ sites
and $M$ down spins:
\begin{eqnarray}
N\theta(x_{\beta})=2\pi J_{\beta}+\sum^M_{{\alpha}=1}
\theta( (x_\beta -x_{\alpha})/2),
\label{x-eq}
\end{eqnarray}
 One sees that the solution for $x_\alpha$ is independent
of both the flux $f$ and on its repulsion $U$.

Let us  substitute
the  variables $\lambda_\beta=U (x_\beta+ \frac{4} {NU}\sum^N_{j=1}
 \sin(k_j))/4$ into
  equation (\ref{momenta}) for the momenta $k_j$ and study the resulting
 equation
in the variables $x_{\beta}$.
Then, analogously to the expansion of  eq.(\ref{lambda}),
expanding the eq.(\ref{momenta}) in powers of $\alpha$ , we get
a new form of the first Bethe equation, namely
\begin{equation}
Lk_j=  2\pi I_j+2\pi f +\sum^M_{\beta=1} \theta(x_{\beta}) +
        \frac {8   \sum^N_{l=1}(1-N\delta_{lj}) \sin k_l } {NU}
\sum^M_{\beta=1} \frac  {1} {1+ x^2_{\beta}}
\label{momenta-1order}
\end{equation}
For parameter $\alpha$  small (dilute density),
this system of linear equations
may be solved, with the result for the momentum
\begin{equation}
k_j= \frac {2\pi} {\tilde{L}} (I_j +f +\frac {1 } {N} \sum^M_{\beta=1} J_\beta
 +\frac {8B } {UL} (\frac {1} {N}\sum^N_{l=1} I_l +f +\frac {1 } {N}
\sum^M_{\beta=1} J_\beta))
                               \label{2momenta-2order}
\end{equation}
and for the ground state energy
\begin{equation}
E_{ground}=- \tilde{D} \cos( \frac {2\pi} {\tilde{L}}
\left(f- \frac {p} {N} +\frac{I_{max}+I_{min}}{2}
+\frac {8B} {UL}  (\frac {1} {N}\sum^N_{l=1} I_l +f -\frac {p} {N}) \right))
 \label {en-corr}
\end{equation}
where $p=- \sum^M_{\beta=1} J_\beta$ , $I_{max}$ and $I_{min}$ are the
maximal and minimal charge quantum numbers,
$$\tilde {D}=2 \sin (\pi N/ \tilde {L})/ \sin (\pi/\tilde {L}),$$
 and
 $$\tilde{L}=L ( 1+ \frac {8B}{UL}).$$
 The magnitude
$$-B=-\sum^M_{\beta=1} \frac  {1}{1+ x^2_{\beta}},$$
 which is a real number,
has the physical meaning of the total energy( ground state + excitations)
of the antiferromagnetic Heisenberg Hamiltonian with
 the exchange constant equal to $1/4$.

The
variables $x_{\alpha}$ are satisfied by the Bethe ansatz equation
 for that Hamiltonian(\ref{x-eq}). Actually,
the value $B(k)$ is a function associated with the
excitations spectrum, where  $k$ is the momentum of the spin-wave
excitation of the Heisenberg Hamiltonian  $k=2 \pi \sum_{\alpha}
 J_{\alpha}/N$. For the antiferromagnetic state of  Heisenberg Hamiltonian
 the dependence $B(k)$ has been studied in detail in many works:
first being  estimated by Anderson\cite{Andsws},
within a spin wave approximation and then
  by  des Cloizeaux and Pearson\cite{Cloi}\cite{Cloi1}
 in the framework of the Bethe ansatz
obtaining the exact form of this spectrum.

At large number $N$ this dependence has the form:
\begin{equation}
B(k)=(N (\log 2-1/4) - \frac{\pi}{2} \mid\sin k \mid)/4
\end{equation}
 The ground state energy(first two terms) has been obtained by
Hulthen \cite{Hult} in the framework of the Bethe method\cite{Beth}.
It was seven years later after  Bethe  introduced his method
(the Bethe ansatz),
applying it to the ferromagnetic Heisenberg chain.

\section{ Single Spin}

Let us consider the case  with $M=1$, i.e. when the system
is strongly polarized.
Here, we find an analytical solution, permitting  eq.(\ref{x-eq})
to be immediately solved yielding
\begin{equation}
x_{\beta}=\tan(\frac{\pi J_{\beta}}{N})
\end{equation}
The function  $B(J_{\beta})$ can then be calculated from
 this solution, giving $B=\cos^2(\frac{\pi J_{\beta}}{N})$.

 Now we must consider two independent cases, when $N$ is odd
and when $N$ is even. Let $N$ be an even number, then $J_{\beta}$
is an integer,  $p$ say. Then,
for the ground state energy we get the explicit expression

\begin{equation}
E_{even-N}=-2 \frac{\sin\big(\frac{\pi N}{L+\frac{8} {U}\cos^2(p\pi/N)}
\big)}
{\sin\big(\frac{\pi}{L+\frac{8} {U}\cos^2(p\pi/N)}\big)}
 \cos \big[ \frac{2 \pi} {L} (f-\frac{p} {N}) \big].
\label{energy-even}
\end{equation}

The ground state energy-flux dependence is this a function
with two types of oscillation.
The period of the first  of oscillation
is equal to the single flux quantum, while the period of
the second type of oscillations is equal to the fraction $1/N$ of the
elementary flux quantum. Let us describe this dependence.
Each
period of the single flux quantum
then
 consists of $N$-parabolic like curves, where each of these curve is
labelled by  $p$: $0\leq p \leq N$.
In the limit of large $N$ we may put as $f=p/N$ to get an exact
single flux periodic function, which has  minima
and maxima  at integer flux values and at half-odd integer flux values
respectively.

For an illustration on Fig.1 is presented the energy-flux dependence
for the Hubbard ring having 1000 sites with 6 electrons and a single down spin
at the coupling $U/t=10$. On this Figure one sees 7 parabolic curves
having  minima at $p/6$ flux values, and 6 cusps associated with
 intersections between each two  neighbour parabolas. The envelope
of these parabolic curves is the single flux periodic
 function $\cos^2 \pi f$.

Next we turn  to the case of $N$ odd, when the numbers $J_{\beta}$ are
half-odd integer. So, let us take $J_{\beta}={ p+ N}/ {2}$, where
 $p$ is an arbitrary integer.
Given this,
then the ground state energy
may be expressed as

\begin{equation}
E_{odd-N}=-2 \frac{\sin\big(\frac{\pi N}{L+\frac{8} {U}\sin^2(p\pi/N)}
\big)}
{\sin\big(\frac{\pi}{L+\frac{8} {U}\sin^2(p\pi/N)}\big)}
 \cos \big[ \frac{2 \pi} {L} (f-\frac{p} {N}) \big],
\label{energy-odd2}
\end{equation}

Again, we find that the ground state energy dependence
 has both the single flux quantum and the fractional
flux quantum periodicity, described by the first and the second
factors, respectively. In the region of the one period
 of the single flux quantum
this dependence again consists of $N$-parabolic like curves,
labelled  by the integer  $p$: $0\leq p \leq N$.
In the limit of large $N$ we may put as $f=p/N$ to get an exact
single flux periodic function. However,
for an
odd number of particles, this function  now
 has maxima at integer flux values
and minima at half-odd integer flux values.

For an illustration on Fig.2 is presented the energy-flux dependence
for the ring having 1000 sites with 11 electrons and a single down spin
at the coupling $U/t=10$. On this Figure one sees 12 parabolic curves
having  minima at $p/11$ flux values, and 11 cusps associated with
 intersections between each two  neighbour parabolas. The envelope
of these parabolic curves is the single flux periodic
 function $\sim-\sin^2 \pi f$.

Several conclusions can be drawn immediately. Firstly,
the parity effect, which disappeared in
the $\alpha$ small limit, is  again recovered
when  the first correction is
taken into account. In other words, the ground state
energy-flux dependence is different for the  even and odd numbers of
fermions on the ring. To be more precise,
 the behavior of the flux-energy dependence for
an even number of particles is shifted by a half flux quantum in comparison
with the case of an odd number of particles.
 This is similar to the parity effect observed on a
 ring with interacting spinless fermions \cite {Legg}, \cite {Kus}
\cite{Loss}.  However, in  comparison
with  spinless fermions on the Hubbard ring the,
 cases of even and odd number of particles are reversed.
 The parity effect discussed might be related to
the number of down-spin electrons, which is $N-1$. In comparison
 with the case of spinless fermions, because of this shifting
of the particle number $N$ by $N-1$,  the
parity effect discussed might be reversed. In this case, again, one may connect
the parity effect with the creation of the statistical flux $\pi$,
due to Fermi statistics\cite{Kus}.

\section{Spin up-down equilibrium}

In this section we  will follow the important suggestion
 by Yu and Fowler \cite {Yu} about
the possible  role
of spin excitations in the Aharonov-Bohm effect on the Hubbard
 ring in the limit of large  $U/t$. Our results
 extend their analysis of the first
corrections in the strong coupling limit as well as
being applicable to systems with the low electron density  at arbitrary
coupling.
Now, let us consider the general case, when the number of  spin-up fermions
is approximately equal to the number of spin-down fermions, i.e. it may differ
by $\pm1$.

   To investigate the behavior of the flux-energy dependence
for such a general situation, we must
consider four independent cases, when the numbers $N$ and $M$ are even
or odd. As we will see below ( also,see, above the single spin case M=1)
for the ground state energy flux dependence
only  the parity of the number $N$ will be important.

 Let us consider the first of these cases, when
$N$  and $M$ are  even. The quantum
numbers (which are half-odd integer)
 of the Heisenberg Hamiltonian, ${J_{\alpha}}$,
for the ground state are distributed symmetrically with respect to
origin, so that $\sum_{\beta} J_{\beta}=0$. The quantum numbers of
the charge degrees of freedom $I_{\alpha}$ for the Hubbard Hamiltonian
are integer with $\sum I_j = -\frac{N}{2} = -M$.

However, at zero
flux the minimum energy of the Hubbard Hamiltonian will correspond to
the set of quantum number $\{J_{\alpha}\}$, but with nonvanishing sum
 $\sum J_{\alpha}=M$, which
 does not correspond to the ground state of the Heisenberg Hamiltonian.
Such a state is associated with the single magnon excitation of the Heisenberg
chain, having a momentum
$k_0 = \frac{2\pi P_0}{N} = \pi$ and  the excitation  energy
$B(k_0) = B_0$. The distribution of  quantum numbers
$\{J_{\alpha}\}$ associated with the ground state of the Hubbard
ring  at zero flux  has a hole near the origin, i.e. is of the form:
\begin{equation}
J_1,\ldots,J_M=-\frac{M-1}{2},\ldots,\frac{-1}{2}, {\bf hole} , \frac{3}{2},
\ldots,\frac{M-1}{2}
\end{equation}
With  increasing  flux    this hole in the distribution  is shifted
to  another place  so that $\sum_{\alpha} J_{\alpha}=M-1$ and, therefore,
 the magnon momentum decreases.  In other words with the flux
the hole in the distribution moves to its left
side. Therefore, with that motion of the hole, exactly
speaking, with the new position of the hole  the magnon
excitation will have the new momentum $k<k_0$ and the energy $B(k)<B_0$.
Exactly speaking, the obtained dependence $B(k)$ describes the low-energy
excitation of the discrete antiferromagnetic Heisenberg chain,
as first discussed  by des Cloiseaux and Pearson\cite{Cloi}\cite{Cloi1}.

With the dependence $B(k)$ taken into account
the appropriate ground
state energy dependence of the Hubbard model on the flux may be written in the
form:
\begin{equation}
E^{even-M}_{even-N}=-\tilde{D} \cos \big[ \frac{2 \pi} {L} (f-\frac{p} {N})
\big],
\label{even-en}
\end{equation}
where the flux  value $f$ is changed in the region $ (2p-1)/2N<f<(2p+1)/N$ and
$\tilde {D}(k)=2 \sin (\pi N/\tilde{L}(k))/ \sin (\pi/\tilde{L}(k))$,
with $k=2\pi p/N$ being the magnon momentum of the finite Heisenberg chain with
$N$ sites, where
$p$ takes integer values.

 The dependence consists of equidistant parabolic segments.
 The position of  each parabolic segment along
the vertical axis depends upon the value $-\tilde{D}(k)$.
As the integer $p$ increases,
the magnon momentum $k$ and
 the energy of this magnon excitation $B(k)$ also increases, so long as
 $p\leq\frac{N}{2}$. The magnon energy
then  has a maximal value at $p=\frac{N}{2}$.

This
means that the value $-\tilde{D}(k)$ associated with
 the position of bottom of the
$p$-th parabolic segment along the vertical axis
 decreases monotonically. As a result, we see
that the envelope curve which passes through the bottom of these
parabolic segments is similar to the  curve describing the
spectrum of the spin-wave excitations of the antiferromagnetic  Heisenberg
chain with $N$ sites. Note, that this curve is
described by the equation $E_{gr}{(k)} = -\tilde{D}(B(k))$.
The flux dependence of this envelope curve can be obtained
if we   substitute $k=2\pi f$.
It is a half
flux quantum periodic function.

    Now, let us consider the second case when {\it $N$ is even, but M is an
odd number}. In this case, at zero flux,
 the ground state energy
corresponds to half-integer numbers $I_j$ and integer numbers
$J_{\alpha}$ distributed  symmetrically  around the origin (see, for
example \cite{Kus}). That is, here,
\begin{equation}
\sum_{j} I_j=\sum_{\beta}J_{\beta}=0
\end{equation}

The analysis of this case is similar to that above.
The given  distribution of quantum numbers
${J_{\alpha}}$ corresponds  to the ground state energy,  both of the
antiferromagnetic Heisenberg chain and the Hubbard ring.
However, including the flux
the ground state energy of the Hubbard ring will now
correspond to the nonvanishing
 sum $\sum_{\beta} J_{\beta}$, which is negative.
This means that there will be
created a magnon excitation similar to that described above,
associated with the appearance of a hole in the distribution of
${J_{\alpha}}$. With the increasing flux the hole will move from
the left  to the right  of the distribution.

We, therefore, once again have the flux-energy dependence of
 the Hubbard ring determined via the low energy
 excitations of the Heisenberg chain.
This dependence again consists of $N$ parabolic segments
and is described by a  similar formula to  eq.(\ref{even-en}). The
envelope, which passes through the bottom of these parabolic segments is
described with the aid of the same equations $-\tilde{D(2\pi f)}$, i.e.
it has cusps at its minima, which are at integer and half-integer
 flux values and smooth
maxima at the values of the flux
exactly midway between the cusps.

    Let us now consider  the case when $N$ is  odd  and $M$ is
 even. The quantum numbers $I_j$ are integer and are
symmetrically distributed around the origin, i.e. $\sum I_j =0$. The
quantum numbers $J_{\alpha}$ are also integers. The ground state energy
of the Heisenberg Hamiltonian corresponds to the dense distribution
with the sum $\sum_{\alpha} J_{\alpha} = -\frac{M}{2}$. However, the
ground state energy of the Hubbard ring with zero flux values must
correspond to the state with $\sum_{\alpha}J_{\alpha}= 0$. This sum can
 equal  zero when the associated state of the Heisenberg
chain is not a ground state, at rather an excited state.
This means
that it is a state with the hole in the $\{J_{\alpha}\}$ distribution. One
sees that in this case the hole will be created exactly
in the origin:
\begin{equation}
J_1, \ldots, J_M :=-\frac{M}{2},\ldots,{-1},{\bf hole},{1},
\ldots,\frac{M}{2}
\end{equation}

This state corresponds to the top in the spin-wave excitations of the
Heisenberg chain, i.e.  to the spin-wave excitation with
momentum $k=\frac{\pi}{2}$. As the flux changes, the parabolic
segment with its minimum at zero flux is exchanged with another the
parabolic segment which has its  minimum at the  flux
$f=\frac{1}{N}$.

 This parabolic segment will correspond to the new set
of $J_{\alpha}$, where the hole is in the position
formerly occupied by
$J_{\alpha} = 1$, so now
 $\sum_{\alpha} J_{\alpha} = -1$.
The appropriate magnon excitation
associated with this parabolic segment will have   momentum
$k=\frac{\pi(N-2)}{2N}$  and   energy $B(\frac{\pi(N-2)}{2N})$,
which are lower
than the  momentum $k=\pi/2$ and the maximal energy
$B(\pi/2)$, respectively.

With the next increasing the flux, there will occur a transition to
a new parabolic segment, corresponding to a  new set of
${J_{\alpha}}$ with the hole at the position $J_{\alpha}= 2$.
    This is the parabolic segment corresponding to the magnon
excitations with  momentum $k=\frac{\pi(N-4)}{2N}$ which has a
lower energy than the magnon excitation in the previous case.
Therefore, at this
  flux the mimimum energy of the appropriate
parabolic segment decreases. The absolute minimum is reached
 at the quater and 3/4  of the flux
quantum, if the flux is changed within a single flux quantum.

 The envelope curve, which goes through the
minima of these parabolic segments is a half flux quantum periodic
function, which has
 smooth maxima at integer and half integer flux values. This curve
may be described with the aid of the formula $E_{ground} =-\tilde{D}(2\pi f)$.

    Let us consider now the last case, which is $N$  and $M$
odd. Here, both the  $J_{\alpha}$  and
$I_j$
are half-odd integer.
 The energy of the Hubbard Hamiltonian,
which corresponds to the dense distributions with $\sum I_j =
-\frac{N}{2}$ and with $\sum J_{\alpha} = \frac{M}{2}$
and assumed to be a ground state
is not a ground state of the
Hubbard ring at zero flux. But this dense distribution of $\{J_{\alpha}\}$ is
the true  ground
state of the Heisenberg Hamiltonian.

The ground state of the Hubbard ring will
correspond to the distribution of the quantum number $\{J_{\alpha}\}$
with the hole at the position $J_{\alpha} = -\frac{1}{2}$ and the sum
$\sum_{\alpha=1}^M J_{\alpha} = +\frac{2M+1}{2} = +\frac{N}{2}$. That
is, to say, in the ground state of the Hubbard ring
there already exists the magnon
excitation with  momentum $k= \frac{2M+1}{N} \pi =
{\pi}/{2}$, which is, in fact, associated with the top in low
energy spin wave excitation spectrum of the Heisenberg chain.
 The energy-flux dependence
 will again consist of parabolic segments
 as described above. The minima
 of these segments will be located at the flux
$f=\frac{p}{N}$. The energy behavior of the minima of these segments
will be exactly the same as described in the previous case.

The picture discussed is illustrated in Figs 3 and 4. In  Fig.3
one sees the coexistence of the half-integer and of the fractional
$1/N$
Aharonov-Bohm effects. The ring has 7500 sites filled with 25 electrons.
There are two types of oscillations present: half-flux quantum periodic,
which has the largest amplitude and the other fractional -
$1/25-$ flux quantum periodic, the amplitude of which
is much smaller.
  The envelope curve of the fractional oscillations is a half
flux quantum periodic function, with  smooth maxima
at integer and half-integer flux
quanta $ f=...,-1/2,0,1/2,1,...$ and cusps, which are minima,
 in between.
This is due to the parity effect, since the number of particles
on the ring is odd.

For  contrast, in Fig.4 we present a  case for an even
 number of electrons  on the ring with
$N=10$, and $L=1000$.
Again,
 as in  Fig.3  the coexistence of the half-integer and fractional
Aharonov-Bohm effects is observed and because there are
 10  particles
the fractional $1/10$ flux quantum oscillations arise.
  The envelope curve of these fractional oscillations
is as in  Fig.3
a half flux quantum periodic function, which, however, has minima at
integer and half-integer flux
quanta.

In the comparing these two Figures, one notes that
with the change of the number of particles from odd to even, the position of
the maxima is shifted by a quarter-flux quantum. This is precisly
due to the parity effect similar to  the parity
effect for spinless fermions\cite{Kus},\cite{Loss},\cite{Legg}.
That is on the Hubbard ring we have obtained the parity effect
which is similar to one for spinless
fermions. The effect  is due to  the
contribution of  the spinon excitation spectrum in the AB effect.
Note, that in the second case (Fig.4) the difference in amplitudes
between the half-integer and fractional $1/N$ oscillations is smaller,
  indicating
that the fractional effect is more important at small number of particles.
With  increasing $U$ or $L$, this difference may disappear.

\section{ Fractional $ {\bf\frac{M}{N}}$ Aharonov-Bohm effect.
         }

In previous sections we have shown that the AB periodicity for the cases
of $M=1$ and $M=N/2$ are different
( single and half flux quantum, recpectively). Therefore, there
 occurs the question: what kind of AB periodicity occurs at
arbitrary $M$: $1<M<N/2$
Let us now consider the case when $M$ is arbitrary, but $M<<N$.
For this general case, when the density of down-spin
  particles $M/N$ is much smaller, then
the density of up-spin  particles $1-M/N$ can again be found
 an analytic solution. In this case we  expand equation (\ref{x-eq})
with in the parameter $M/N$. We can assume that
the values of $x_{\beta}$ in eq.(\ref{x-eq}) are small. Using
these assumptions
the eq. (\ref{x-eq}) may be written in the form
\begin{equation}
x_{\beta}+\sum^M_{\alpha=1} \frac{x_{\alpha}-x_{\beta}}{2 N} =\frac{\pi
J_{\beta}} {N}
\end{equation}
i.e. we have obtained a system of linear equations, which can  be
immediately solved, giving
\begin{equation}
x_{\beta}=\frac{2 \pi J_{\beta}} {N-2 M}-\sum^M_{\alpha=1} \frac{\pi
J_{\alpha}}{(2 N -M) N}
\end{equation}
This solution looks like  the momentum spectrum of a free particle on
a chain of $2N-M$ sites ( the first term of the right hand side of the
equation), plus an additional fictitious flux associated with
the second term of the right hand side of
the equation. From the solution  obtained
 one sees that the values $x_{\beta}$
are small if $\pi M/(2N-M)<<1$, which is equivalent to our
small parameter $M/N<<1$, and therefore $x_\alpha<<1$
does, indeed, hold.

On the other hand this analysis has shown that
the second term of the right hand side of  eq.(\ref{x-eq}) is smaller
that the first one in the parameter $\frac{M}{N}\ll 1$.
Therefore, in
that case the solution may be obtained in the form:
\begin{equation}
x_{\alpha} = \tan \frac{\pi J_{\beta}}{N}
\end{equation}
The substitution of which into $B[x_{\alpha}]$ gives
\begin{equation}
B(k) = \frac{M}{2} + \frac{1}{2}  \sum_{\beta=1}^M
\cos \frac{2\pi J_{\beta}}{N}
\label{eq-B}
\end{equation}
The choice of the set ${J_{\alpha}}$ determines the dependence of the
low lying excitations $B(k)$. Now we must consider four independent
cases associated with the parity of numbers $N$ and $M$.

    Let, first us consider the case, when $N$ is even and $M$ is odd. Then
the ground state energy corresponds to the vanishing sums of the
integers $J_{\alpha}$,  $\sum_{\alpha} J_{\alpha}=0$ and of the
half-integers $I_j, \sum_j I_j=0$. Both of these sets are densly
distributed, symmetrically around the origin. The excitation will
correspond to the hole in the distribution of ${J_{\alpha}}$. For
example, when $p=1=-\sum J_{\alpha}$, then the hole is created
on the left hand side, from
the left edge
. For the value $p=2$, the hole is moved on
to the next position
. In general, if the hole in the position
$J_{\alpha}$ then  we can write

\begin{equation}
p= -\sum_{\beta=1}^{M+1} J_{\beta} + J_{\alpha}
\label{mom-1}
\end{equation}
where  the sum is taken over the dense distribution. Therefore, if
 $p\leq M$, this sum equals $\sum_{\beta=1}^{M+1} J_\beta = -(\frac{M+1}{2})$.
In
the similar way, by splitting into to terms,
 the sum over the dense distribution of
$M+1$ quantum numbers $J_\beta$ and the term associated with the
hole at the place of $J_\alpha$  we find for the energy of such excitations:
\begin{equation}
B(k) = \frac{1}{2} (\frac{M}{2} +
\frac{sin\frac{\pi(M+1)}{N}}{sin\frac{\pi}{N}} cos\frac{\pi}{N} - cos
\frac{2\pi J_{\alpha}}{N})
\end{equation}
Then, expressing $J_{\alpha}$ from eq.(\ref{mom-1})
and substituting it into eq.(\ref{eq-B}) we get the dependence
of $B(\frac{2\pi p}{N})=B(k)$ where $p\leq M$. Note, that the case of
$p=-\sum_{\beta=1}^{M} J_{\beta}=M$
 corresponds to the ground state distribution of $J_{\alpha}$ shifted
by 1 to its left side. When $p>M$, in this shifted distribution there
 will again
appear the hole, initially,  at the beginning on the left hand side,
 from the left edge.

    With  increasing $p$ this hole will move from the
left to the right side, once $M<p\leq 2 M$. At the value $p=M$, the
distribution of $J_{\alpha}$ will be dense, but shifted by 2 to the left
in  comparison with the distribution associated with the value
$p=0$. Thus, for an arbitrary value of $p$, we can write

\begin{equation}
p= + l (M+1) + (\frac{M+1}{2}) - J_{\alpha}
\label {mom-gen}
\end{equation}

The energy of the associated excitations is equal to:

\begin{equation}
B(p) = \frac{1}{2} [M+\frac{sin\frac{\pi(M+1)}{N}}{sin\frac{\pi}{N}}
cos(\frac{2\pi l + \pi}{N}) - cos \frac{2\pi}{N} (p - \frac{M+1}{2}  -l(M+1))]
\label{gse-even,odd}
\end{equation}

where $l=0$, if $0\leq p \leq M $,
 $l=1$, if $M\leq p\leq 2M$,   $l=2$, if
$2M\leq p\leq 3M$ and so on.

    Thus, the ground state energy has cuspoidal minima (cusps) at $p=nM$,
where $n$ is an integer. So, if we substitute $p=lM$, we have for
the envelope functions of these cusps:
\begin{equation}
B(p) = \frac{1}{2} (M + \frac{sin(\frac{\pi M}{N})}{sin\frac{\pi}{N}}
cos\frac{2\pi l}{N})
\end{equation}
 Since $p\leq\frac{N}{2}$, the function $B(l)$ is monotonically
 decreasing with $l$, having a maximal value when
$l\sim[\frac{N}{2M}]$, where the brackets indicate that we take the integer
part of the value $\frac{N}{2M}$. The spectrum of the excitations
of the Heisenberg chain reflects the flux quantum periodicity of the
Hubbard ring, as described above. Therefore, we come to the conclusion,
 that in addition to fractional ${1}/{N}$, and integer  on the Hubbard
ring there also occurs fractional $\frac{M}{N}$ flux quantum periodicity.

For an illustration we give several examples for the ring with
 $U/t=10$ and $L=10^4$,
presented in Fig.5. In the first example (see, Fig.5a) the energy-flux
dependence is given for
$N=20$ and $M=3$ in the region of the half flux quantum. Since
the number of particles is not large one sees, the very pronounced
$1/N$ oscillations associated with the single parabolic like segment.
However, one sees already $M/N$ oscillations with much smaller
amplitude. So, the first, fourth and seventh parabolic like segments
have minima lower than nighbouring parabolas. If the number
of particles increases, for example, $N=30$ (see, Fig.5b, where we
 taken $M=5$) the amplitude
of $M/N$ oscillations increases. In this case the amplitudes of $1/N$
and are $M/N$ oscillations are, already, compared.
With the next increasing number of particles $N$, for example,
when $N=40$ and $M=7$ presented in Fig.5c, the amplitude
of $M/N$ oscillation becomes larger than the amplitude of the
$1/N$ oscillation.

    Let us consider the second case, when $N$ is odd and $M$ is even,
which corresponds to two types of integer quantum numbers
${J_{\alpha}}$ and ${I_j}$, with $\sum I_j=0$ and
$\sum_{\alpha}J_{\alpha}= -\frac{M}{2}$. However, the ground state of
the Hubbard ring will correspond to an excited state of the Heisenberg
chain with the hole in the origin, i.e. the ground state
 corresponds to the sum:
$\sum_{\alpha} J_{\alpha}=0$.

    The excitation associated with the value $p=1$ corresponds to the
shift of this hole to the position $J_{\alpha}=1$. So with the
increasing $p$ the hole moves to right.
Therefore, we can
write $p= -\sum_{\alpha=1}^{M+1}J_{\alpha} + J_{\beta}=J_{\beta}$,
where $p\leq {M}/{2}$ and the sum is taken over the dense
distribution.

 As $p$ is  increased  from
${M}/{2}$ the hole will be created on the left edge and moved to
the right edge, where $p\leq(M+{M}/{2})$. With the next increase
of $p$ this process will be repeated and so on. Thus, we may write in
general form that

\begin{equation}
p= +l(M+1) + J_{\beta}  ,
\end{equation}
where the first term related to the sum $\sum^{M+1}_{\alpha=1} J_{\alpha}$.

    The envelope function
of parabolic like curves
associated with the ground state energy-flux dependence
 is calculated in a similar way and
equals:
\begin{equation}
B(p) = \frac{1}{2} [M+ \frac{sin\frac{\pi(M+1)}{N}}{sin\frac{\pi}{N}}
cos(\frac{2\pi l}{N}) - cos\frac{2\pi}{N}(p-l(M+1))]
\label{gse-odd,even}.
\end{equation}
where  $l=0$, if $-(M)/2\leq p \leq (M)/2$; $l=1$, if
 $(M)/2<p\leq  (M)/2+M $,   $l=2$, if
$ (M)/2+M\leq p\leq  (M)/2+2M$ and so on.

 This dependence is shifted by $\frac{M}{2N}$ is comparison with the
previous case. It consists of cuspoidal minima and smooth maxima in
midway between.
 In comparison with the previous case
here the cuspoidal minima, one of which
existed at the  value $p=0$, are shifted by the same value of the flux,
too, and are exchanged in the positions with
smooth maxima. Thus,
 the properties of the envelope function are the same as
in the previous case.
For an illustration in Fig.6 we present the energy-flux
dependence for the ring with $U/t=10$. So, it is clearly
coexist $6/19 \sim 1/3$ and $1/19$
flux quantum periodicities for the ring with 19 electrons having 6 down-spins
for number of sites $L=1000$
(see, Fig.6a). The parabolic like curves are associated
with $1/N$ type oscillation.
 The envelope function of these curves is $6/19 \sim 1/3$ flux quantum
periodic function. The cusps of this function are located at $3/19$ and
$9/19$ flux quantum. With the increasing number of particles
the amplitudes of both type of fractional oscillations decrease.
This can be seen in Fig.6b, where the particles number was changed to $49$
and the number of sites was taken as $10^4$.
Although with the next increase
of the number of particles  at the fixed value of $M$
the amplitudes of
both type of oscillation decrease,
the fractional $1/N$ oscillations begin
to wash out, however the fractional $M/N$
oscillations are still exist. This
is shown in Fig.6c where the number of particles
was taken as $N=99$ on the ring with the same the number of sites,
as in Fig.6b. On the other hand when both numbers, the number of
particles $N$ and the number of dowm-spins $M$ increase,
the amplitude of $M/N$ oscillation increases and it is compared
with the amplitude of the main single flux quantum
periodic AB oscillation. In the latter the fractional $1/N$ oscillation
is completely washed out, although it is an origin for the fractional
$M/N$ oscillation. The fractional $23/99$ AB
oscillation is presented
in Fig.6d, where we consisder the ring with $M=23$ and $N=99$.
One sees on this Figure
that the fractional $1/99$ oscillation is practically
washed out.

The other two cases only differ from these already considered
ones only
by an appropriate shift.
In both of them the ground state of the Hubbard ring corresponds to the
excited state of the Heisenberg chain with the value
$$p_0=-\sum_{\alpha=1}^{M}J_{\alpha} = -\frac{N}{2} .$$

Because of the
$M$-periodicity, this state may be mapped into another one,
 where the sum
$ \sum^M_{\alpha=1} \tilde{J_{\alpha}}=0$ vanishes, i.e.
 this can be obtained with the aid of the transformation:
$$
\tilde{J_{\alpha}}=J_{\alpha}-l_1,
$$
where $l_1$ is the integer part of
the ratio $\frac{N}{2M}$. Therefore,
 we can write
$$
\sum_\alpha\tilde{J_{\alpha}}= \sum_{\alpha}J_{\alpha}-l_1 M=q ,
\label{eq-q}
$$
 where $q<[\frac{M}{2}]$.

Thus,
 the problem has been reduced to the one, discussed above, but with the
difference, that the ground state of the Hubbard ring will correspond
to the excited state of the Heisenberg chain where the hole in the
dense distribution of the set ${J_{\alpha}}$ is located at the $-q$-th
position, which is determined
by eq.(\ref {eq-q}). The other excited state
of the Heisenberg chain (or the ground state energy-flux dependence
of the Hubbard ring)
 is associated with the shift of this hole
to the right edge. Then all excitations may be described exactly by the
same manner as above. That is we may write
\begin{equation}
\tilde{p}= -\sum_{\alpha=1, dense}^{M+1}\tilde{J_{\alpha}} + \tilde{J_{\beta}}
\label{m30}
 \end{equation}

  Now we must consider the two particular cases ( 1) $N$ and $M$ are
even and 2) $N$ and $M$ are odd), independently.

 In the first case, the
values $J_{\alpha}$ or $\tilde{J_{\alpha}}$ are half-odd integer and from
eq.(\ref{m30} ) we get
\begin{equation}
\tilde{p}= \frac{M+1}{2} + \tilde{J_{\beta}}
\label{mom-3}.
\end{equation}
    The excitation energy is estimated as above yielding:
\begin{equation}
B(p) = \frac{1}{2} [ M+ \frac{sin\frac{\pi(M+1)}{N}}{sin\frac{\pi}{N}} cos
(\frac{2\pi l + \pi}{N}) - cos\frac{2\pi\tilde{J_{\beta}}}{N}]
\label{energy-3} .
\end{equation}

{}From
 eq.(\ref{mom-3}) and making the inverse
transformation from
$\tilde{J_{\alpha}}$ to $J_{\alpha}$ we finally get
\begin{equation}
B(p) = \frac{1}{2}( M+ \frac{sin\frac{\pi(M+1)}{N}}{
\sin\frac{\pi}{N}}
\cos(\frac{2\pi l+\pi}{N}) -
\cos\frac{2\pi}{N} [p+ \frac{N}{2} - \frac{M+1}{2} -l(M+1)])
\label{gse-even,even}
\end{equation}
where the value $l$ is choosen thus
$l=0$, if $-N/2\leq p \leq -N/2+M$;
 $l=1$, if $-N/2+M\leq p\leq  -N/2+2M$,
 $l=2$, if $-N/2+2M\leq p\leq -N/2+3M$ and so on.

For an illustration we present in Fig.7 the dependence of the
ground state energy for the ring with 1000 sites and $U/t=10$.
The number of electrons is taken as $N=20$ amd $M=6$. One sees  on this
Figure that the dependence of the envelope function of the parabolic curves
is monotonically decreased with the flux increases from zero.
This dependence has cusps, which are repeated with the periodicity
$6/20$. So we have here a quasi $\sim 1/3$ flux quantum periodicity.

    Similarly way we can obtain the excitation energy for the second
case, when the parity of  particles number is odd, i.e.
 $N$ and $M$ are odd:
\begin{equation}
B(p) = \frac{1}{2} [ M+ \frac{sin\frac{\pi(M+1)}{N}}{sin\frac{\pi}{N}}
cos\frac{2\pi l}{N} - cos\frac{2\pi}{N} [ p+\frac{N}{2} - l(M+1)]]
\label{gse-odd,odd}
\end{equation}
where the value $l$ is chosen
thus
 $l=0$, if $-N/2-(M-1)/2 \leq p \leq -N/2+(M-1)/2$;
 $l=1$, if $-N/2+(M-1)/2 \leq p\leq  -N/2+(3M-1)/2$,
 $l=2$, if $-N/2+(3M-1)/2\leq p\leq -N/2+(5M-1)/2$ and so on.

For an illustration we present in Fig.8 the dependence of the
ground state energy for the ring with 1000 sites and $U/t=10$.
The number of electrons is taken as $N=19$ and $M=5$. One sees  on this
Figure that the dependence of the envelope function of the parabolic curves
is monotonically decreased with the flux increases from zero.
This dependence has cusps, which are repeated with the periodicity
$5/19$. So, here again as in the previous case
 we have  a quasi-$\sim 1/3$- flux quantum periodicity.

    Thus, for all four cases the excitation energy is as ${M}/{N}$
periodic function, having cusps repeated with the
same periodicity. To study the flux dependence we must
 substitute the flux value $f$ instead of
the value $p/N$ of the excitations of Heisenberg ring.
 Generally speaking, in the
last two cases  the ground state  energy-flux dependence
of the Hubbard ring related to these excitations of the Heisenberg
chain is shifted by half-flux quantum in comparison with two
previous cases, respectively. But, the qualitative
 shape of these dependencies are not changed.

\section{Fine Structure}

    One general conclusion that can be drawn here that
the complete picture of the Aharonov-Bohm effect must include
a fine structure. The latter appears when we take into
account the electron-electron interactions and is characterized
by oscillations with an amplitude and a period smaller than
the conventional one. In fact, there is a hierarchy of  flux periodicities
which depends implicitly on the Zeeman interaction.

The Fourier spectrum
of the energy-flux
dependence for the Hubbard ring mainly consists of two or three harmonics.
The number and type of these harmonics depend on the contribution
from the Zeeman interaction.
 One of these harmonics is conventional and
has a single flux quantum period; the second one has the fractional
${1}/{N}$ flux quantum period; the third one has half-flux
quantum period or fractional $M/N$ period. The latter appears
 when the number of
down spin particles $M$  and
the number of up-spin particles
$N-M$ are different.

Let us analyse the fine structire.
The amplitude of the single-flux or the fractional $M/N$
periodic oscillations is proportional to the parameter $ n^2 \alpha$,
where $n={\pi N}/{L}$ and we use dimensionless units.
The amplitude of the fractional $M/N$ oscillation
is smaller, by factor $M/N$, than the amplitude of
oscillations with the integer period. Here, we assume that $M<<N$.
When $M \sim N/2$ the situation is reversed and the $M/N$ oscillation has
a dominant amplitude.
 The amplitude
of the fractional ${1}/{N}$ flux quantum periodic oscillations
 is proportional to
$\frac{n^2}{ N^3}$ and does not depend on the value $U$.
One sees also that both types of oscillations only depend
upon $U$ and $L$ through
  their product $UL$ or in other words
they only depend upon $\alpha$. With  increasing  $N$ at
 constant $n$ and $\alpha$,
the amplitude of the  fractional $1/N$ oscillations vanishes.

    The  single flux quantum harmonic coexisted with fractional $M/N$
 periodic harmonics show the parity
dependence. That is there is a difference in the energy-flux
dependencies for the cases of an  even and an odd number of fermions. This is
in  contrast with the  fractional ${1}/{N}$ periodicity, which has no the
parity
dependence. In the present case the parity effect
for an even and an odd number of fermions is transposed
  with respect to
 the case of
 spinless
fermions.

    It is useful explicitly  to write  an envelope of the
flux-dependence. For example in the case, where there is up spin- down spin
equilibrium (i.e. M=N/2),
 for the two cases of an even or an odd number of particles on the ring
the flux energy dependence has the form:

\begin{equation}
E_{erv} = \frac{n^3}{3} \frac{t^2}{U}
\left\{
\begin{array}{ll}
        \mid \sin 2\pi f \mid &\mbox { $N$ is even}\\
       \mid \cos 2\pi f\mid &\mbox{ $N$ is odd}
       \end{array}
\right\} .
\end{equation}

For the other values of $M$ the prefactor will be the same,
but the oscillation factor will be changed to reflect the fractional
$M/N$ flux quantum periodicity.

 On the
other hand the amplitude of the fractional ${1}/{N}$ oscillations
is estimated to be $\sim {tn^2}/{N^3}$, which is rapidly
 decreases  with
increasing $N$. Comparing this amplitude
 with the amplitude of $E_{erv}$ we get the
criterion for when the single flux quantum or $M/N$ flux
quantum  periodicity becomes dominant over the fractional $1/N$ ones;
 this
happens when
\begin{equation}
N\gg \sqrt[4]{\frac{UL}{t}}
\end{equation}
    However, to investigate this problem we have used the parameter
$\alpha = \frac{Nt}{UL} \ll 1$. The region where
the single or $M/N$ flux quantum periodicity is dominant is
\begin{equation}
\sqrt[4]{\frac{UL}{t}} \ll N\ll \frac{UL}{t}
\end{equation}
One sees that at large $L$ the limits when such behavior exists may be
easily recovered.

    In conclusion, then in the dilute density limit or for
 strong coupling
 we have discovered two
different regimes of the Aharonov-Bohm effect, namely, fractional $1/N$ and
integer coexisting with fractional $M/N$ and $1/N$. Integer and fractional
$M/N$ periodic oscillations appear simultaneously and may have
comparable amplitudes, which are distinct by numerical factor.
As a rule, in high magnetic fields
 the amplitude of $M/N$ oscillations is always amaller
than the amplitude of integer ones.

Thus, a very surprising implicit role of the Zeeman interaction has been
found, i.e.  with  increasing  magnetic field, as the
contribution of the Zeeman energy increases, in addition to
the fractional $1/N$ flux quantum periodicity, there appears the
integer-fractional $M/N$ flux quantum periodicity. With  increasing
 the magnetic field, i.e. with  decreasing $M$
the half flux  quantum periodicity at $M=N/2$ transforms
 into fractional $M/N$ ones and integer one.
Finally, in strong field, when the system is almost
polarized, only the integer flux quantum periodicity remains.

This transformation is not continuous but singular. The period
of this fine structure of the AB effect changes
with each  spin flip.
 For almost polarized system, we have
the coexistence of integer  and fractional $1/N$ flux quantum periodic
oscillations, only.
 On the ring with a small number of electrons
$N\leq\sqrt[4]{\frac{UL}{t}}$, there occurs
the fractional ${1}/{N}$ Aharonov-Bohm effect.
 However with  increasing  $N$, when
$N\gg \sqrt[4]{\frac{UL}{t}}$,
the single-fractional $M/N$ flux quantum
periodicity becomes dominant over the fractional $1/N$ effect,
although they  may both coexist.

\section{               CHANGE OF THE FINE STRUCTURE PERIOD
                         WITH MAGNETIC FIELD}

    The period of fractional oscillations
$\frac{M}{N}$ related to the number of down-spin particles. However,
 with the increase of magnetic field on the ring there may
occur spin-flip processes. As the result the number of the down-spin
particles changes (decreases) and the magnetization of the ring
related to the number $M$ decreases. This Zeeman magnetization may be
described with the aid of the formula:

\begin{equation}
(N-2M) = [\frac{2\chi H}{\mu_B}]
\end{equation}
where brackets means an integer part of the value.
The value $\chi$ is a
susceptibility of the Hubbard chain . First, in thermodynamic limit the
value of $\chi$ has been calculated by Takahashi \cite{Taka} for the case
of the half-filling, then for an arbitrary filling the value of $\chi$
has been calculated by Shiba \cite{Shib}. For the low density limit
$\alpha<<1$ we discuss the value of $\chi$ may be estimated by the similar
way. For the large values of $L$ the thermodynamic limit is
acceptable. Therefore we  use Shiba's result

\begin{equation}
\chi = \frac {3 U  L} { 2 \pi^4 t^2 n^2}
\end{equation}
whence
\begin{equation}
M = \frac{N}{2} - [\frac{\chi H}{\mu_B}]
\end{equation}
    One sees that with the field the value of $M$, and, therefore,
 the period of the
fine structure decreases
\begin{equation}
f_T = \frac{1}{2} - [\frac{\chi H}{\mu_B}]
\frac{1}{N}.
\end{equation}

    Thus we came to the conclusion that in the Aharonov-Bohm effect of
strongly correlated system such as the Hubbard ring may occur the
oscillation of the smallest amplitude, which period decreases, when
the field increases. In such an effect the spin orbit interaction
might be important (see, for example, Ref.\cite {Fuji})

\section { towards reality from exact solutions}

In the present work we have presented the exact solution
for the Hubbard ring, showing
a fine structure in the Aharonov-Bohm effect.
The mesoscopic rings in existing experiments have many channels,
large sizes and many electrons ( see, also, discussion in Ref.\cite{Avis}).
They also have disorder or impurities. The measurements are
carried at finite temperatures. Of course, all these factors make
a large difference between the presented model calculations and
the realistic situations.

However  on such  rings the electron density is much smaller
than in a 3D metal, where  screening exists.
Therefore  the electron-electron
interaction is very important.
The effects of the strong
electron-electron correlations, which give rise
the finite structure of AB effect, of course, will appear there.
 But at present time there is no
tool to take into account  correctly the  many-body effects related
to this interaction. The
Bethe ansatz method is only one single tool, where the effects of strong
correlations may be studied for large number of electrons
and  a system of a large size, but for some models only.
So, using the Bethe ansatz  we found, that
the interaction between particles also may create the finite
structure, which is characterized by several types of
periodic oscillations with the amplitude and period smaller than
one unit.

    In the integer-fractional regime the effects of
 disorder and  temperature will be to make
the flux-energy dependence smoother leading to
the disappearence of the fine structure. In experimental situations, where
the disorder or the temperature always exist, but have a low level
for the ring with large number of
electrons one may see three types of oscillations
in the flux energy dependence. With the next increase of the disorder
or the temperature the oscillations having the smallest
amplitude will disappear. So, for example, instead of integer-fractional $M/N$
periodic oscillations
on the single ring, one may see only the integer flux quantum
periodic oscillations.

 The discovered single flux quantum periodicity, when the finite structure
has disappeared, may be relevant to
two recent experimental studies of the Aharonov-Bohm effect on
single metallic and semiconductor rings\cite{21},\cite{22}.
Recent experiments\cite{21} have described only the full quantum period in
the case of individual metallic rings. On a semiconductor single
 loop in the $GaAs/GaAlAs$ system \cite {33}
such single flux periodicity has been detected.
To explain this integer flux  quantum periodicity we assume  that
 the individual rings are sufficiently disordered. Recently,
 similar  results\cite{Weis} \cite{Gogo}
were obtained that free electrons on
the single ring with disorder or with the temperature may show
 only the full single flux quantum periodicity.
Both these findings are consistent with   existing experiments.

The interaction between particles  may also support the full
single quantum periodicity, and, therefore, it may  also be
responsible for the observed single flux quantum period.
The prediction of fractional $1/N$ and fractional $M/N$
oscillations could be made for the ring
having lowest level of disorder. Once the resolution of the experimental
studies has been improved such periodicities  may be observed.

The other experiment\cite{22}, which is worth to mention
is that the half-flux quantum periodicity simultaneously of $10^7$ rings
 has been measured. It is
 also consistent with our results.
It is reasonable to assume that  there is
 a disorder. So the single ring must show
the single flux periodicity. For such large ensemble of
the rings we must take also into account the parity effect. Then
it is also reasonable that half of these rings has an even number
 of particles and the other half of these rings has an odd number of particles.
Because of the discovered  and described parity effect,
averaging over the two cases of odd and even number of particles indicated
above, immediately gives the $\Phi_0/2$
period. This interpretation is then natural for the experimental results
concerning many rings\cite{22}.
Thus in the framework of our approach we have obtained the unified
pictures why the
$\Phi_0$ period is seen in individual rings and why $\Phi_0/2$ period
is seen an ensemble of  $10^7$ rings.

In addition to these explanations, we predict the existence of
oscillations with  periods smaller than $\Phi_0$
or  $\Phi_0/2$.
To check the predictions of our theory,
 it is important
to perform the Fourier analysis of the persistent current oscillations
 in order
to detect the fractional Aharonov-Bohm effect, which
coexists with integer one, i.e.
whether there are  fractional oscillations on a background of
integer ones.

 Recently the fractional $1/4-$ AB effect
has been observed
\cite{Liu} in $AuIn$-rings prepeared
by $e-$beam lithography .
Since this effect has been observed in the region of a
superconducting phase transition,
 where the phase separation occurs. It seems that
there the effective system
reminding the quantum dot chain ring is created. Therefore,
the discussed Hubbard model for the description of
unpaired electrons in these droplets may be applicable.

In this case  the fractional $1/4-$ effect  may be created
as a result of averaging over rings
 having different parity of electron's number.
This occurs only at a small magnetic field
when Zeeman energy may be neglected. With the increasing magnetic field
the electrons become partially polarized and
such periodicity is destroyed. But here the fractional $M/N$ may appear
instead.

 Thus, our
finding of a coexistence of the fractional $1/N$ effect with fractional
$1/4$ or $1/2$  integer AB effects may not only to explain the observed
$1/4$ flux quantum periodicity but also predicts smallest, namely
$1/N$ and $M/N$ fractional AB effects, once the experimental
resolution has been improved.

\vspace{3mm}
\begin{flushleft}
{\bf Acknowledgement}
\end{flushleft}
\vspace{2mm}

We thank  T. Ando, S. Katsumoto, M. Kohmoto,
M. Ueda, K.  Kawarabayashi, F. Assaad
and M. Kohno for useful discussions of obtained results
and especially S.M. Manning for careful reading of the manuscript
and corrections.
 The work by FVK  has been supported by
 Ministery of Education, Science and Culture of Japan, he also appreciates
the hospitality of ISSP.

permanent address:

{\parskip=0pt
$*)$L.D. Landau Institute for Theoretical Physics \par
Moscow,117940, GSP-1, Kosygina 2, V-334, Russia \par }

\newpage
{\large \bf Figure Captions}\\

\bigskip

{\bf Fig.1}  The behavior of the ground state
 energy  as a function of
flux  $f$ for 6 electrons with the single up spin at the values $L=1000$ and
$U=10$ in the region
of flux within the single fundamental flux quantum.
The energy is expressed in
the units $t  10^5$ . The zero energy corresponds to $-11.9993 t$.

\bigskip

{\bf Fig.2}  The behavior of the ground state
 energy  as a function of
flux  $f$ for 11 electrons at the values $L=1000$ and $U=10$ in the region
of flux within the single fundamental flux quantum. One electron has down spin
i.e. $M=1$. The zero energy corresponds to $-21.995 t$.

\bigskip

{\bf Fig.3}  The behavior of the ground state
 energy  as a function of
flux  $f$ for 25 electrons at the values $L=7500$ and $U=10$ in the region
of flux within the single fundamental flux quantum. 12 particles have
up-spin and 13 particles have down-spin.The energy is expressed in the
units $t  10^6$ . The zero energy corresponds to $-24.999544 t$.

\bigskip

{\bf Fig.4}  The behavior of the ground state
 energy  as a function of
flux for 10 electrons at the values $L=1000$ and $U=10$ in the region
of flux within the single fundamental flux quantum. Five particles have
up-spin and five particles have down-spin.
 The energy is expressed in the units $t  10^5$ .
 The zero energy corresponds to $-9.9983 t$.

\bigskip

{\bf Fig.5}  The behavior of the ground state
 energy  as a function of
flux for the ring with even number of electrons $N$ and with
odd number for projection of angular momentum
$M$, for the parameters $L=10^4$ and $U=10$.
We present the
 cases when  $N$ and  $M$ are equal to:
 a) $N=20$ and $M=3$ , energy is expressed in arbitrary units,
 the region of flux  is the half of fundamental flux quantum.
b)  $N=30$ and $M=5$ , energy in    units $t  10^6$,
the zero energy corresponds to $ -59.99911 t$; the region
of flux  is the half of fundamental flux quantum.
c)  $N=40$ and $M=7$ , energy is expressed in the units $t  10^7$,
the zero energy corresponds to $ -79.997896  t$; the region
of flux  is the half of fundamental flux quantum.

{\bf Fig.6}  The behavior of the ground state
 energy  as a function of
flux for the ring with odd number of electrons $N$ and with
even number for projection of angular momentum
$M$, for the parameters $L=10^4$ and $U=10$.
We present the
 cases when  $N$ and  $M$ are equal to:
 a) $N=19$ and $M=6$ , energy is expressed in the units $t  10^6$,
the zero energy corresponds to $-37.97752 t$; the region
of flux  is the half of fundamental flux quantum.
b)  $N=49$ and $M=6$ , energy is expressed in the units $t  10^7$,
the zero energy corresponds to $-97.996131t$; the region
of flux  is the half of fundamental flux quantum.
c)  $N=99$ and $M=6$ , energy is expressed in the units $t  10^6$,
the zero energy corresponds to $-197.96808t$; the region
of flux  is the quarter of fundamental flux quantum.
d)  $N=99$ and $M=23$ , energy is expressed in the units $t  10^6$,
the zero energy corresponds to $-197.96809t$; the region
of flux  is the half of fundamental flux quantum.

{\bf Fig.7}  The behavior of the ground state
 energy  as a function of
flux for the ring with even number of electrons $N$ and with even
number of a projection of the angular momentum
$M$, for the parameters $L=10^3$ and $U=10$.
We present the case when $N=20$ and $M=6$,
 energy is expressed in the units $t  10^6$,
the zero energy corresponds to $-39.97377 t$; the region
of flux  is the half of fundamental flux quantum.

{\bf Fig.8}  The behavior of the ground state
 energy  as a function of
flux for the ring with odd number of electrons $N$ and projection of angular
momentum
$M$, for the parameters $L=10^4$ and $U=10$.
We present the case when $N=19$ and $M=5$,
energy is expressed in the units $t  10^6$,
the zero energy corresponds to $-39.97752 t$; the region
of flux  is the half of fundamental flux quantum.

\end{document}